\documentclass[twocolumn,trackchanges]{aastex63}
\submitjournal{AJ}

\usepackage{graphicx}
\usepackage{dcolumn}
\usepackage{bm}

\usepackage{makecell}
\usepackage{verbatim} 
\usepackage{amsmath}

\begin{document}


\title{Magnetic Effects and 3D Structure in Theoretical High-Resolution Transmission Spectra of Ultrahot Jupiters: the Case of  WASP-76b }

\correspondingauthor{Hayley Beltz}
 \email{hbeltz@umich.edu}

\author[0000-0002-6980-052X]{Hayley Beltz}

 \affiliation{Department of Astronomy, University of Michigan, Ann Arbor, MI 48109, USA}

 \author[0000-0003-3963-9672]{Emily Rauscher}
\affiliation{Department of Astronomy, University of Michigan, Ann Arbor, MI 48109, USA}

\author[0000-0002-1337-9051]{Eliza M.-R.\ Kempton}
\affil{Department of Astronomy, University of Maryland, College Park, MD 20742, USA} 
 
 \author[0000-0003-0217-3880]{Isaac Malsky}

 \affiliation{Department of Astronomy, University of Michigan, Ann Arbor, MI 48109, USA}
 
\author[ 0000-0002-2454-768X]{Arjun B. Savel}
\affil{Department of Astronomy, University of Maryland, College Park, MD 20742, USA}

\begin{abstract}
High resolution spectroscopy has allowed for unprecedented levels of atmospheric characterization, especially for the hottest gas giant exoplanets known as ultrahot Jupiters (UHJs). High-resolution spectra are sensitive to 3D effects, making complex 3D atmospheric models important for interpreting data. Moreover, these planets are expected to host magnetic fields that will shape their resulting atmospheric circulation patterns, but little modeling work has been done to investigate these effects. In this paper, we generate high-resolution transmission spectra from General Circulation Models for the canonical UHJ WASP-76b with three different magnetic treatments in order to understand the influence of magnetic forces on the circulation.  In general, spectra from all models have increasingly blueshifted net Doppler shifts as transit progresses, but we find that the differing temperature and wind fields in the upper atmospheres of these models result in measurable differences. We find that magnetic effects may be contributing to the unusual trends previously seen in transmission for this planet. 
Our $B=3$ Gauss active drag model in particular shows unique trends not found in the models with simpler or no magnetic effects. 
The net Doppler shifts are additionally influenced by the dominant opacity sources in each wavelength range considered, as each species probes different regions of the atmosphere and  are sensitive to spatial differences in the circulation. This work highlights the ongoing need for models of planets in this temperature regime to consider both 3D and magnetic effects when interpreting high resolution transmission spectra. 
\end{abstract}

\section{Introduction}
High resolution spectroscopy (HRS, typically R $\gtrsim$ 30,000) has opened windows into exoplanet atmospheres at an unprecedented level of precision. HRS has allowed for detections of new atmospheric species as well as measurements of net Doppler shifts and broadening due to atmospheric winds and rotation \citep{Snellen2010,Brogi2016,Schwarz2016}. \cite{Brogibook2021} offers a recent review of the techniques and major results of HRS.

Due to the level of precision offered by HRS (and the fact that planets \textit{are} multi-dimensional objects), three-dimensional (3D) atmospheric models are ideal for interpreting these spectra. Previous work has found that not only do 3D effects show up in high resolution transmission \citep{Louden2015,Flowers2019} and emission \citep{Herman2022,vansluijs2022,Pino2022} spectra, but detection strengths can increase when using spectra generated from a 3D model compared to 1D models \citep{Beltz2021}. Typically the 3D structure of exoplanet atmospheres are simulated with General Circulation Models (GCMs). This type of numerical model solves the set of fluid dynamical equations known as the ``primitive equations of meteorology" to simulate a planet's atmospheric structure, its including temperature and wind fields throughout its orbit. 

Ultrahot Jupiters (UHJs)  are ideal for testing differing treatments of magnetic effects. Due to the thermal ionization of dayside species \citep{parmentier2018,Helling2021}, charged particles will be blown around the planet and interact with magnetic field lines generated from the planet's interior dynamo \citep{Perna2010magdrag}. Partially due to their already significant computational time, most GCMs do not have an explicit treatment for magnetic effects in their simulated atmospheres. One commonly used treatment is applying a global uniform Rayleigh drag timescale to the atmosphere such as in GCMs from \cite{Tan_2019,Carone2020,Deitrick_2020THOR,Lee2022}. Notably this timescale is also sometimes used to encompass a variety of effects, also including  hydrodynamical ones. Although easy to numerically implement, this prescription of magnetic drag includes assumptions that become problematic when applied to planets with strong day-night temperature differences. Since the strength of magnetic effects is a strong function of local ionization levels (and so temperature), order of magnitude estimates of the global field strength corresponding to a particular uniform drag timescale \citep[such as those carried out in][]{Kreidberg2018} effectively imply that the global magnetic field is nearly two orders of magnitude stronger on the nightside than the dayside, for the case of the UHJ WASP-76b and assuming a $10^{4}$s uniform timescale \citep{Beltz2022a}. Instead, for a uniform global magnetic field, we should expect magnetic effects to be much stronger on the dayside, compared to the negligibly ionized nightside \citep{Perna2010magdrag,Beltz2022a}. 
The most physically consistent treatment of magnetic effects are found in specialized non-ideal magnetohydrodynamic (MHD) models \citep[such as those presented in][]{Rogers_2014_showman, Rogers_2014b,Rogers2017}, but at the cost of simplifying other aspects of the modeling such as the treatment of radiative transfer\citep[see][for a more detailed discussion]{Beltz2022a} , and greatly increased computational time required, resulting in less than a handful of these types of models having been published for UHJs. In this work, our models use a medium complexity ``kinematic MHD" approach, allowing the strength of the drag timescale to vary as a function of temperature, pressure, and latitude. \citep[For a more detailed description of this approach, see][]{RauscherMenou2013,Beltz2022a}.

UHJs are the ideal laboratory for exploration with high-resolution spectroscopy due to their extremely favorable signal to noise ratio caused by their size and temperature. Here, we specifically focus on the UHJ WASP-76b, an inflated gas giant orbiting an F-type star with a period of 1.81 days \citep{West2016}. High resolution transmission spectra for this planet have been observed and studied extensively \citep{Cassayas2021,Deibert2021,Landman2021,Taberno2021} with a recent work \citep{Kesseli2022w76spectralsurvey} exploring the wide range of species detected in the atmosphere of the planet. An influential transmission result by \cite{Ehrenreich2020} found a spatially asymmetric and extremely large blueshift ($-$11 km/s) of neutral iron, arguing this blueshift is a result of nightside condensation of the species. Alternate physical processes have been suggested to explain this large blueshift, including clouds, non-zero eccentricity \citep{savel2022no}, or large temperature differences between limbs \citep{Wardenier2021}, but so far, it has been difficult for GCMs to match this magnitude of shift. A recent work by \cite{Gandhi2022} performs a deep analysis on this dataset, providing constraints on both temperature and Fe abundances for four different regions of the planet and confirming spatial differences across the terminator.

 In this work, we explore modeled high-resolution transmission spectra for three different models of the UHJ WASP-76b: one with a Uniform drag timescale, one with our kinematic MHD approach, and one without any treatments for magnetic effects. The difference in temperature and wind structures of these three different models result in spectra that vary throughout transit, opening the door to the exploration of how magnetic effects can alter high resolution transmission spectroscopy. This work represents the first time the impact of magnetic drag assumptions on high resolution transmission spectra has been studied \citep[we similarly explored the impact of magnetic effects on high resolution emission spectra in][]{Beltz2022b}. By identifying measurable differences between high resolution transmission spectra simulated using different prescriptions for magnetic effects, we can hope to predict how we might empirically constrain the role of magnetism in UHJ atmospheres. 

In Section \ref{sec: Methods}, we briefly describe the models used in this analysis and the different treatments for magnetic effects we tested. We also discuss our radiative transfer post-processing and choice of wavelength ranges to generate our predictive spectra. In Section \ref{sec: Results}, we explore the features of our predicted spectra and examine the impact of magnetic model and wavelength choices. We then put this work in context of the model's assumptions and other capabilities in Section \ref{sec:Discussion}. Finally, we summarize our main conclusions in Section \ref{sec:Conclusion}.

\section{Methods} \label{sec: Methods}
\subsection{GCM}
For this work, we post-process previously generated 3D GCMs of the ultrahot Jupiter WASP-76b  \citep[first published in][where more details and specific numerical parameters can be found]{Beltz2022a} using a ray-striking radiative transfer code to generate high-resolution transmission spectra at multiple wavelengths and resolutions.  These models used the RM-GCM \citep{Rauscher2012GCM,newradRomanRausher} with parameters appropriate for WASP-76b, with 65 vertical layers evenly spaced in log pressure, from 100 to $10^{-5}$ bars, and a horizontal spectral resolution of T31, corresponding to roughly $\sim 4$ degree spacing at the equator. The simulations ran for a total of 2000 planetary days. Our GCM assumes hydrostatic equilibrium, which is a valid assumption for the opacity sources included in the high-resolution spectra we calculate from these models. This is relevant to note as recent work from  \cite{Zhang2022Hydrostatic} finds that absorption strength of particular species often detected in transmission of UHJs, such as FeII and H$\alpha$ can't be explained from hydrostatic equilibrium assumptions. This is not an issue for our work due to our choices of opacity sources.

The models from \cite{Beltz2022a} were calculated for several different magnetic drag prescriptions at a variety of field strengths; we choose to analyze the same subset of models as in we did in \cite{Beltz2022b}. These models differ in the way they treat magnetic effects, as follows: 
\begin{itemize}
    \item Drag Free/0 G: This is the baseline model that contains no additional forms of drag to represent magnetic effects. The GCM does contain numerical hyperdissipation and three sponge layers \citep[see][for a discussion on sponge layers in GCMs]{Beltz2022a}, both of which are used for numerical stability and are also present in the models listed below.  
    
    \item Uniform/$10^{4}$ s: This method of applying drag is often found in GCMs due to its numerical simplicity. A single Rayleigh drag timescale---in this case $10^{4}$ seconds---is applied throughout the simulation to the horizontal and vertical momentum equation. This value was chosen to match the strong drag case from \citep{Tan_2019} and provide the same comparisons as the analysis work presented in \citet{Beltz2022b}.
    
    \item Active drag/3 G: This method for treating magnetic effects, first used in \citet{RauscherMenou2013} and first applied to UHJs in \citet{Beltz2022a} is the most physically complex treatment of magnetic effects that we test. Our active drag prescription, also sometimes referred to as a ``Kinematic MHD'' treatment, also applies a drag on the winds, but only in the east-west direction \citep[as geometrically appropriate for a dipole global field][]{Perna2010magdrag} and with a timescale calculated based on local conditions, using the following expression from \citet{Perna2010magdrag}:
\begin{equation} \label{tdrag}
    \tau_{mag}(B,\rho,T, \phi) = \frac{4 \pi \rho \ \eta (\rho, T)}{B^{2} |sin(\phi) | }
\end{equation}
where $B$ is the chosen global magnetic field strength (in this case 3~G), $\phi$ is the latitude, $\rho$ is the density,  and the magnetic resistivity ($\eta$) is calculated in the same way as \citet{Menou_2012}: 
\begin{equation} \label{resistivity}
    \eta = 230 \sqrt{T} / x_{e} \textnormal{ cm$^{2}$ s$^{-1}$}
\end{equation} 
where the ionization fraction, $x_{e}$, is calculated from the Saha equation, taking into account the first ionization potential of all elements from hydrogen to nickel \citep[as in][]{RauscherMenou2013}.
\end{itemize}
There are currently no direct observational constraints on the magnetic strength of this planet, or any exoplanet for that matter.  Although in \citet{Beltz2022a} we present a variety of active drag field strengths (0.3~G, 3~G, and 30~G), we are primarily focusing on the 3~G model, as it is the best match for previously published \textit{Spitzer} phase curves from \cite{May2021} \citep[as shown in][]{Beltz2022a}. We previously found that varying the magnetic field strength changed how deep the magnetic circulation regime---characterized by dayside flow up and over the poles---persisted. All of these models exhibit this magnetic circulation at the high pressures probed by high-resolution transmission spectroscopy. Thus, we chose the 3~G model as a representative for the active drag models. This field strength is also in line with estimates from interior modeling by \citet{Yadav2017}.

It is important to acknowledge that our GCM currently does not consider H$_{2}$ dissociation and recombination. This process is expected to reduce the day-night temperature contrast \citep{Bell_2018,Pluriel2020} of UHJs. Another important result of dissociation is the change in scale heights throughout the atmosphere. On the dayside, the mean molecular weight is decreased due to the dissociation, thus increasing the scale height. However, at the same time, the temperature of this region is decreased, meaning a potential reduction in scale height. The nightside wouuld show the opposite trend (increasing in temperature and mean molecular weight). Recent work from \citet{Savel2023}  explores the effect of scale height differences in limb asymmetry during transit. Future work should explore the interaction between this process and magnetic effects.  

\begin{figure*}
    \centering
    \includegraphics[width=\linewidth]{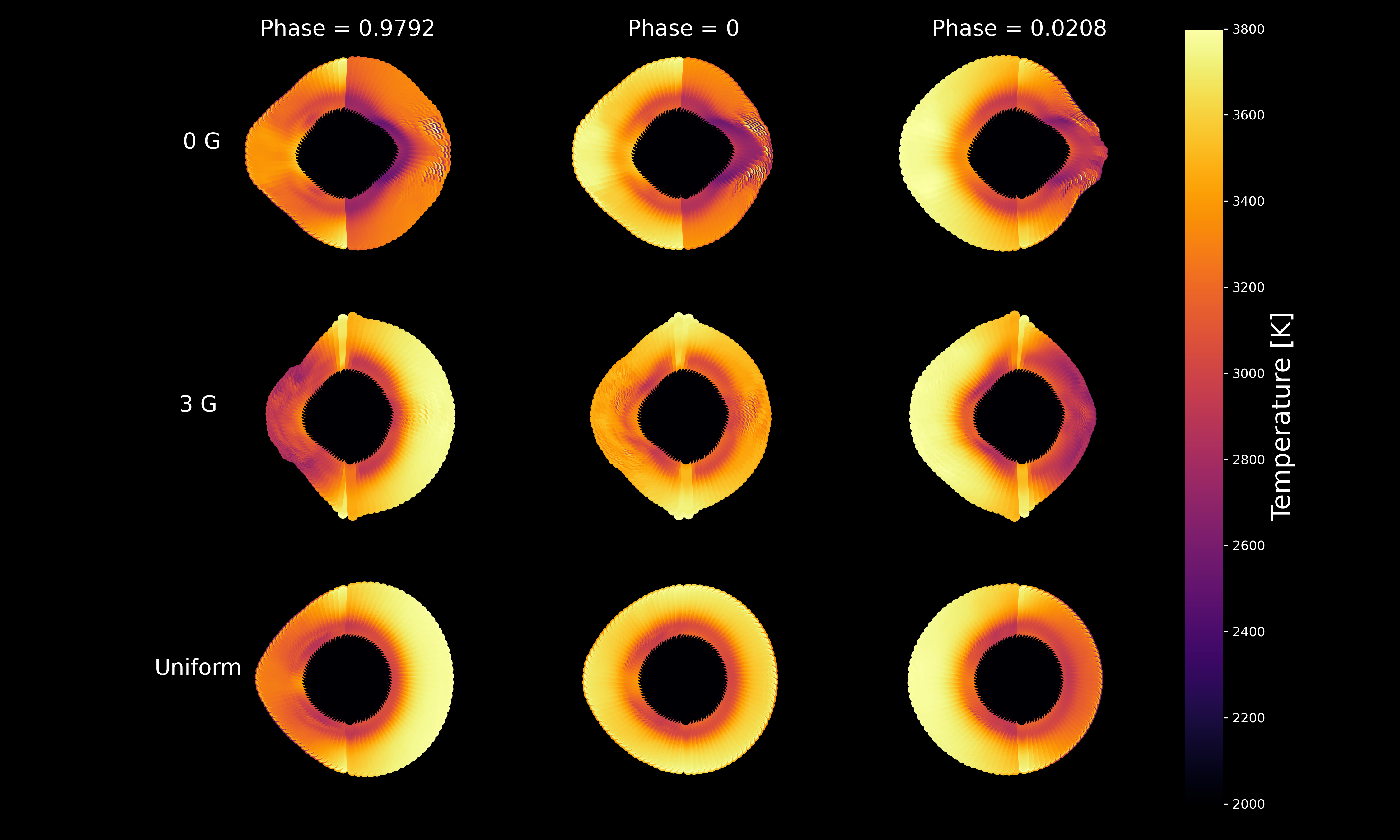}
    \caption{Temperature projections  for the three models used in this analysis, for pressures less than $\sim$0.1 bars. Note that the core and atmosphere are not to scale, but the relative altitudes at different locations are accurately plotted. The east and west limb asymmetries in spatial extent is a result of the difference in scale heights of each region due to non-uniform temperatures between the east and west terminators. Because of the planet's short orbital period, it rotates considerably ($ > 30^{o} $) between ingress to egress, which is reflected above.     } 
    \label{fig: tempmap}
\end{figure*}

The temperature distributions of the models, for the near-terminator regions probed by transmission spectroscopy, are shown in Figure \ref{fig: tempmap}, which plots the temperature structure at a slice of the planet as it would be oriented during ingress, mid-transit, and egress. Note that only the upper atmosphere (maximum pressure of $\sim$0.1 bars) is shown and the relative size between the atmosphere and planet core is not to scale. From this plot, we can see that the spatial vertical extent of each limb varies throughout transit, with the hotter regions being much more extended. The 3~G active drag model shows the most variation across the limbs at ingress and egress but at transit center the 0~G model shows the strongest temperature variation.   This is directly related to the fact that the 3 G model has the largest day-night temperature contrast of the models considered and that this planet rotates $ > 30^{o} $ throughout the entire transit. 

\begin{figure*}
    \centering
    \includegraphics[width=\linewidth]{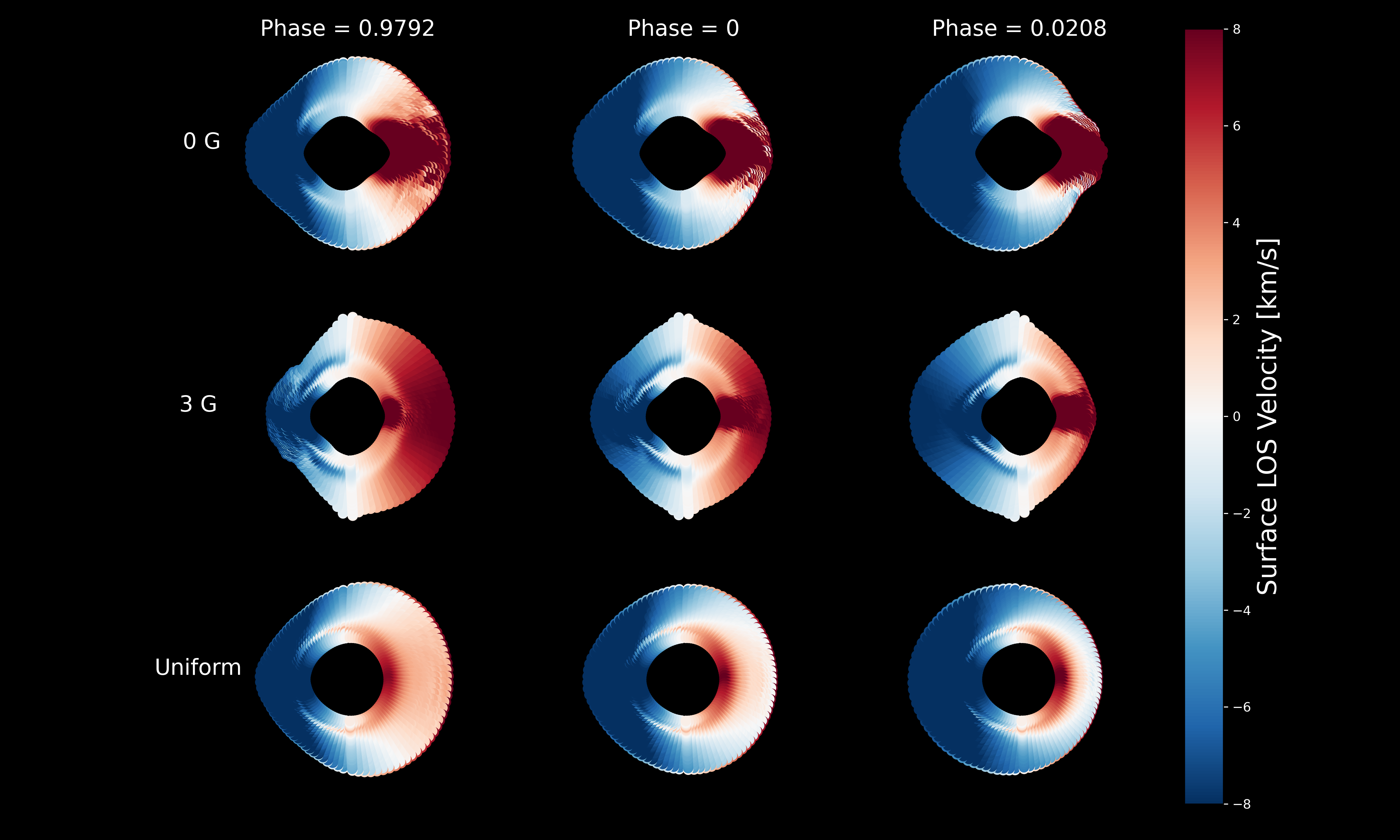}
    \caption{  Line of sight velocities for the three models considered in this work at ingress, mid-transit, and egress. Throughout transit, the blueshifts dominate the net Doppler shift for all models, though the magnitudes of the net Doppler shifts are both model and wavelength dependent.   } 
    \label{fig: losvel}
\end{figure*}

It is also important to consider the line of sight velocities due to strong winds of each model, as shown in Figure \ref{fig: losvel}. As the transit proceeds, the planet rotates, allowing different parts of the atmosphere and their associated winds to come into view. These winds will directly influence the net Doppler shifts associated with each model. We delve deeper into these calculations in section \ref{sec: Results}, but by eye one can notice that the 3~G model has the strongest redshifted regions of the three and the Uniform drag has the strongest blueshifted regions. One can additionally see  that the 0~G model displays some high-altitude and high-latitude winds that are blowing in the substellar-to-antistellar direction. But, since that direction includes an east-west component, this flow structure is disrupted in the 3~G active magnetic drag model and so that blue-shifted contribution to the net Doppler shift is removed. 

\subsection{Radiative Transfer}
We use the same method of calculating high resolution transmission spectra that accounts for 3D effects as that described in detail in \citet{Kempton2012,savel2022no}. In short, the output from our GCM (containing temperature values, east-west wind speeds, and north-south wind speeds at every grid point) is interpolated onto a constant altitude grid so that the post-processing radiative transfer can consistently implement line-of-sight ray striking that calculates intensity and then transit depths at each wavelength. During this process, winds from the GCM and the planet's bulk rotation are incorporated via Doppler shifts in the local opacities. Stellar limb darkening effects are accounted for, meaning that as the planet progresses through transit, the projected stellar flux illuminating each region of the planet's atmosphere is adjusted based on  the limb-darkening coefficients found in \citet{Ehrenreich2020}. 

\subsubsection{Calculated Transmission Spectra}
We calculate high resolution transmission spectra from our three models for three different wavelength ranges, each with a different opacity source  of interest:
\begin{itemize}
    \item Wavelength 1: 0.379-0.789  $\mu$m; R=400,000; Opacity source: Fe 
    \item Wavelength 2: 1.135-1.355 $\mu$m; R=125,000; Opacity source: H$_{2}$O
    \item Wavelength 3: 2.3-2.35 $\mu$m; R=125,000; Opacity source: CO

\end{itemize}
The latter two wavelength ranges match the work done in \citet{Beltz2022b}. The opacity sources of particular interest are noted above, but both sets contain opacity from the following six species: CO, H$_{2}$O, TiO, VO, K, and Na. Relative abundances of these species were calculated assuming solar-abundance \citep{Lodders2003} equilibrium models with \texttt{FastChem} \citep{Fastchem2018,Stock2018}. Wavelength range 2 covers a range accessible by multiple high-resolution spectrographs including WINERED \citep[][]{WINERED2016} and CARMENES \citep[][]{CARMENES2014}. Wavelength range 3 overlaps with the IGRINS instrument \citep[][]{IGRINS2014}. Both of these wavelength ranges are probed by the CRIRES+ instrument \citep[][]{CRIRES+2014}.
The first wavelength range matches the observations of this planet taken by the ESPRESSO spectrograph, first published in \citet{Ehrenreich2020}. This set is also unique in that the only included source of opacity is Fe.

The choice of opacity sources is motivated by theory and observational results. To start, each of these three species are expected to absorb strongly in their corresponding wavelength range \citep{kurucz1995kurucz,rothman2010hitemp,polyansky2018exomol,Fastchem2018}.  Fe was chosen to allow a direct comparison to the data presented in \citet{Ehrenreich2020}. We chose CO due to its expected near uniform abundance \citep[as shown in Figure 1 of ][]{Beltz2022b}. Additionally, recent work from \citet{Savel2023} suggests CO represents an ideal tracer molecule for UHJ atmospheres, given this expected uniformity in abundance in these atmospheres. Finally, we chose to examine H$_{2}$O due to its lack of uniformity in abundance \citep{parmentier2018}. The daysides of UHJ are hot enough to disassociate water, reducing its abundance on the hotter limb. The net Doppler shifts resulting from this dissociation provides an interesting comparison to those from the CO spectra.  
All spectra were calculated assuming local thermochemical equilibrium and solar abundances. Recent work from \cite{Gandhi2022} suggests a metallicity for this planet slightly higher than solar, but consistent with solar within error bars presented.    

\subsubsection{Spatial Distribution of Opacity Sources}
The wavelength regimes that we produced spectra for were chosen partly because of the differing main opacity source. Given the extreme temperature contrasts of the planet, our opacity sources are not necessarily uniformly distributed around the planet. We will briefly touch on the spatial distribution of the main absorbers for each wavelength here. 
\begin{itemize}
    \item Wavelength 1, Fe: Fe is expected to have a non-uniform abundance distribution in the atmosphere. For cooler regions of the planet, Fe is expected to condense, potentially into optically thick clouds. \footnote{Notably, these models were ran without active clouds, so Fe condensation is applied in the radiative transfer post-processing.} Work from \cite{savel2022no} suggests there would be more Fe on the eastern limb of the planet. Additionally, \cite{Wardenier2021} found a lack of gaseous iron on the western limb allows for the signal from \cite{Ehrenreich2020} to be reproduced. 
    \item Wavelength 2, H$_{2}$O: Given that the dayside is hot enough to thermally disassociate water, which is accounted for in the radiative transfer, the abundance of water between the morning and evening terminators differs by roughly 3 orders of magnitude for the 0~G model, but less than one order of magnitude for the other models.   
    \item Wavelength 3, CO: Given the extremely strong triple bond of this molecule, even UHJ atmospheres will not dissociate this species \citep{parmentier2018,Savel2023}. Additionally, the night side is warm enough such that CO is not expected to convert into methane. Thus, its global distribution is essentially uniform the planet.

\end{itemize}

\section{Results} \label{sec: Results}

\begin{figure}
    \centering
    \includegraphics[width=3.5in]{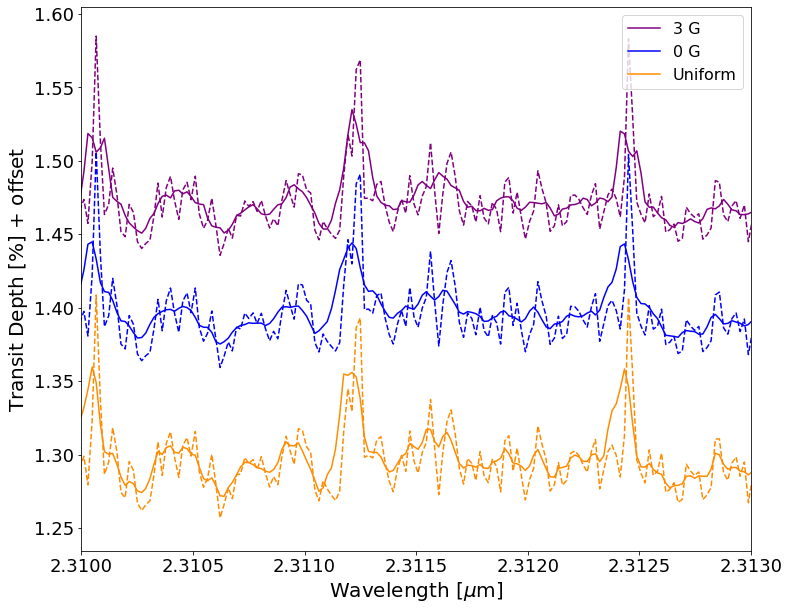}
    \caption{Mid transit (phase=0) spectra from our three models. The solid lines show spectra that have been shaped by Doppler shifts due to winds and rotation while the dotted curves do not have this influence. Vertical offsets have been added for clarity. Differences between the spectral features are due to the different temperature and wind patterns of the models. All three models have similar vertical temperature structures, and so the spectra without Doppler effects are only very subtly different, while the different wind patterns between the models result in noticeable differences in the resulting spectra. } 
    \label{fig: midtransitallmodel}
\end{figure}

We begin our analysis by first searching for differences in the spectra by eye. In Figure \ref{fig: midtransitallmodel} we show the calculated transmission spectra at mid-transit (phase=0) for all three models from a subset of wavelength range 3 (2.3-2.35 $\mu$m) for versions of the spectra calculated with and without Doppler effects from winds and rotation.  The spectra without Doppler effects are difficult to differentiate by eye as the differences between the spectra are on order of $0.5 \%$, but these small differences are a result of differing temperature structures between the models.   We more clearly see the differences between the models in the broadened spectra, as these models have unique upper atmosphere wind structures due to the different types of drag applied to each model.  

\begin{figure}
    \centering
    \includegraphics[width=3.5in]{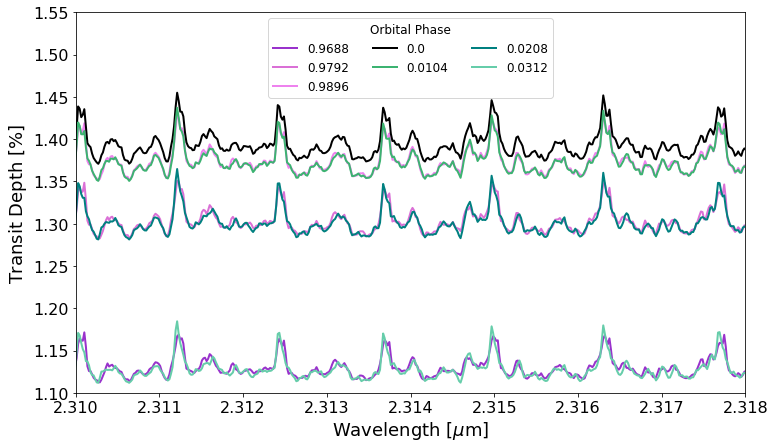}
    \caption{Simulated high resolution transmission spectra from our 3~G model, with Doppler effects from winds and rotation, shown at equally spaced times throughout transit. As expected, mid transit (phase=0) has the strongest absorption since the maximum amount of light is obscured by the planet's atmosphere at this phase. The first and last phases shown here are partial transits, which is why their continuum values are lower compared to the other spectra shown. We can compare spectra that are equally spaced in time before and after mid-transit to identify differences due to east-west asymmetries around the terminator. While the phases nearest to mid-transit are very similar, those phases further away show larger differences, with the spectra near the end of transit (where only a portion of the planet is transiting the host star) showing very sharply blue-shifted lines. } 
    \label{fig: 3Gallphase}
\end{figure}

Because WASP-76b rotates significantly during transit, its spectra will vary as different parts of the atmosphere come into view \citep{Gandhi2022,Wardenier2022}.  In Figure \ref{fig: 3Gallphase}, we show transmission spectra produced from the 3~G model, where Doppler shifts and stellar limb darkening have been applied. Since the spectra are evenly spaced in phase from mid-transit (phase=0), the spectra appear in ``pairs'' where spectra sharing  the same absolute offset from transit have similar continuum levels. These pairs are not identical though; differences in line center (due to differing wind patterns) and absorption strength (due to differing temperature structure) exist. It is also noticeable that during the second half of transit, lines become more blueshifted as the more spatially extended side of the planet increasingly dominates the back-illuminated part of the planet's atmosphere. Similar trends were found in \cite{savel2022no} and \cite{Wardenier2021}, in line with the trend presented in \cite{Ehrenreich2020}.  Thus, the 3D geometry of the model is making a noticeable difference in the resulting high resolution transmission spectra. 

\begin{figure*}
    \centering
    \includegraphics[width=6.5in]{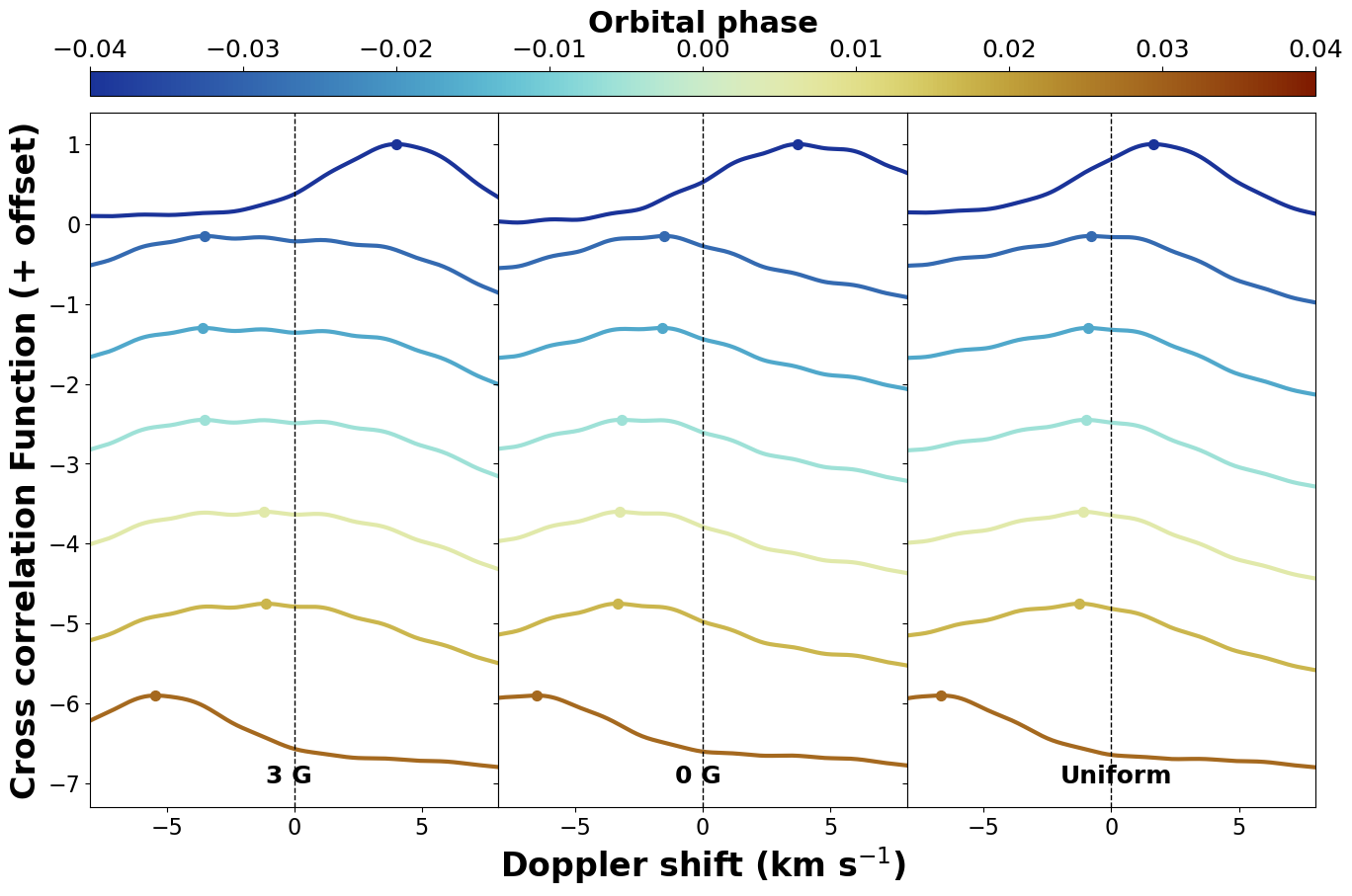}
    \caption{ Cross correlation curves for each model throughout transit (with the first and last points being partial ingress and egress respectively) for wavelength range 1 (1.135-1.355 $\mu$m). The peak of the cross correlation curve, corresponding to the net Doppler shift of the spectrum are shown with colored points. These net Doppler shifts vary with time and differ between models due to the differing circulation patterns. Notably, for this wavelength range, the 3~G spectrum become \textit{less} blueshifted for a time near the end of transit. This is a unique feature of the 3~G model and is a result of the differing wind structure caused by the active magnetic drag prescription.   } 
    \label{fig: transmissioncc}
\end{figure*}

A tool often used in high-resolution spectral analysis is that of cross-correlation between the data and a template spectrum (in velocity space), which we can use to combine the information from all of the lines in a spectrum. If the Doppler-on version of a spectrum is cross correlated with the corresponding Doppler-off version of the same spectrum, one can determine the net Doppler shift at that phase by finding the corresponding velocity of the peak of the cross correlation function, as shown by the points on the curves in Figure \ref{fig: transmissioncc}, calculated for wavelength range 2 (1.135-1.355 $\mu$m). Broadly, our models show a changing net Doppler shift becoming more blueshifted with time, due to the increasing contribution from the more extended, hotter eastern limb, whose motions from winds and rotation are oriented toward the observer during transit. Interestingly, the 3~G models are an exception to this for a short time after mid-transit where the net Doppler shift becomes less blueshifted for a brief time before becoming more blueshifted by the end of transit. It is also relevant to note that of all the cross-correlation curves presented in this Figure, the 3~G curves are the most broadened and least peaked, particularly near mid transit. The broadness of these curves can be attributed to the dual existence of  strong blueshifted and redshifted winds in the upper atmosphere.

\begin{figure*}
    \centering
    \includegraphics[width=1.05\linewidth]{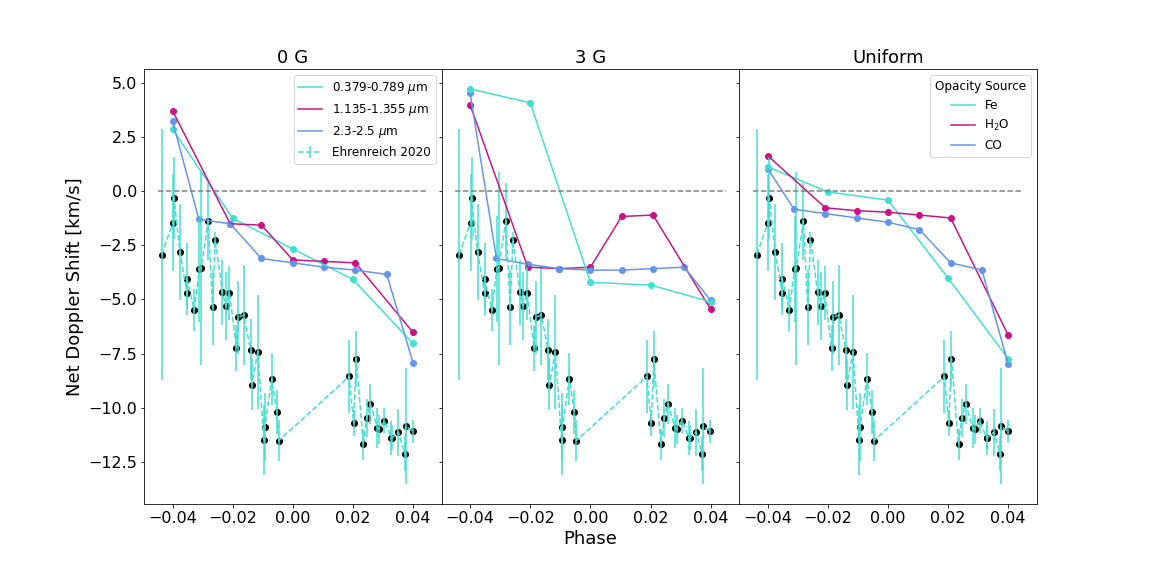}
    \caption{Net Doppler shifts for the simulated spectra from each model, over the three wavelength regimes considered, as well as the data from \cite{Ehrenreich2020}. Overall, the net behavior shows the spectra becoming more blueshifted as transit proceeds. However, the 3~G active drag model shows some deviations from this shortly after mid-transit, but only when water is the dominant absorber within the wavelength range considered. When iron is the dominant absorber, we see instead that the starting net redshift persists longer into transit, before switching to a net blueshift.  Both of these behaviors are unique to the 3~G model, distinguishing it from the 0~G or Uniform drag models.} 
    \label{fig: dopshifts}
\end{figure*}

Figure \ref{fig: dopshifts} shows the net Doppler shifts for all the models considered in this work at each wavelength range. Each model exhibits unique trends but overall, the spectra become more blueshifted throughout transit. The Uniform drag model consistently shows the strongest blueshifts at each phase examined. Similarly, the 0~G model wavelength becomes more blueshifted throughout transit for each wavelength range. The 3~G model shows interesting structure in each wavelength regime, with the spectra covering the near-IR becoming slightly less blueshifted right after mid-transit.

The differences in these net Doppler shifts between models can be attributed to a variety of physical effects. First, the underlying velocity structure of the upper atmosphere differs between each model. For example, in \cite{Beltz2022a}, we saw significantly different dayside wind structures for the active drag model, with the dayside winds traveling up and over the poles in the North-South direction. The influence of these differing flow patterns can be seen in Figure \ref{fig: losvel}. Additionally, differences in relative scale heights due to atmospheric temperature differences will affect the net Doppler shifts \citep{Wardenier2022,Savel2023}. Figure \ref{fig: tempmap} shows this, with the hottest atmospheric regions having the largest vertical extent.

The spatial extent of the dominant opacity source (determined by the wavelength) will influence the net Doppler shift \citep{savel2022no}.  We can start by examining the wavelength range containing CO, as this species is fairly uniform in abundance around the planet. The 0~G and Uniform drag models show similar behavior for this wavelength range, but with the Uniform drag having the strongest net blueshift. We can attribute this to the weakness of the redshifted region on the western limb, which is stronger for both of the other models.  For the 3~G case, we see that the net Doppler shift is roughly constant, aside from the first and last phases calculated. This means that although more blueshifted regions are coming into view during transit, this effect is roughly equaled out by the redshifted winds on the western limb and the different scale heights associated with each limb.

Water on the other hand is certainly not uniformly distributed around the planet. This is most easily seen in the case of the  wavelength range 2 for the 3~G model, which actually becomes less blueshifted with time for part of egress.  Water will dissociate in hot temperatures, so it is less abundant in blueshifted limb. However, as transit progresses, we see a water-depleted blue limb and a relatively water-rich red limb, resulting in a brief period during transit where the spectra becomes more redshifted. Since this effect is not seen to the same degree in the other two wavelength ranges tested, one can infer the feature is influenced by the dominant absorber, water. Although temperature inhomogeneities between the limbs exist for all models presented, the 3~G model has the strongest day-night temperature contrast and most dominant red-shifted atmospheric winds. This particular combination of atmospheric structures results in the behavior seen in Figure \ref{fig: dopshifts}.\textit{ Neither the Uniform or 0~G models show this behavior, indicating that these net Doppler shifts may be a way of testing approximations of active drag.} 

Fe abundance is slightly more temperature dependent than CO, but not nearly to the same level as water. While water abundances can very by over 6 orders of magnitude from the dayside to the nightside of this planet, Fe abundances only change by less than a single order of magnitude, and is slightly more abundance in the cooler regions of the planet. 

For  wavelength range 1 where Fe is the dominant opacity source, , we can also make a direct comparison to the Doppler shifts measured in \cite{Ehrenreich2020}.  
Although the data displays stronger magnitudes of blueshifted values than our models predict, the magnitude of this shift can be altered strongly by small changes in orbital parameters \citep{savel2022no}, leading us to instead  focus on comparing
the velocity trends with orbital phase between the models and the data. The 3~G model does the best job of reproducing the trend found in the data. Both experience a strong negative slope in Doppler shift shortly before mid transit and roughly constant Doppler shifts throughout the rest of transit. The Uniform and 0~G model have a roughly constant slope which does not match the data as well.  Thus, out of the different drag prescriptions tested, our active drag model best matched the trend presented in \cite{Ehrenreich2020}. Notably, other GCM work has struggled to reproduce this trend---particularly the``bottoming out" or ``kink" behavior that occurs after mid transit. \citet{Wardenier2021}  removed iron from the leading limb of their atmosphere to reproduce this ``kink'' while \cite{savel2022no} used optically thick clouds and a slight non-zero eccentricity to best fit the data. However, the models from these two works also incorporated uniform drag in their atmospheres. Thus, this interesting behavior in the \cite{Ehrenreich2020} dataset may be a result of some combination of magnetic effects, clouds, or Fe condensation, however we refrain from making more detailed predictions until a model with both of these effects in concert is presented, which we leave for future work.

\section{Discussion} \label{sec:Discussion}

While we have presented particular features in transmission spectra of UHJs that could be used to assess drag mechanisms within the planet's atmosphere, it is important to recognize the necessary limitations of our modeling and any potential impact this could have on our results.

One caveat to this work is that due to numerical stability purposes, the top boundary of our model is $\sim 10^{-5}$ bars. Compared to emission observations, transmission spectra probe a higher region of the atmosphere, potentially at lower pressure values than what is contained in our model. These regions are less dense with potentially stronger wind speeds. This upper boundary could be contributing to why our net Doppler shifts are not as large in value as those reported in \cite{Ehrenreich2020}, although other GCMs similarly struggle to produce such large shifts \citep[][despite these GCMs covering nearly 2 orders of magnitude more in pressure space]{Wardenier2021,savel2022no}. 

Additionally, the GCMs studied in this work use a double-gray radiative transfer scheme instead of a more complex picket-fence or correlated-k method. A downside of the double-gray method is that it results in more isothermal upper atmospheres than the other radiative transfer schemes mentioned \citep[][]{Lee2021}. This effect is minimized on transmission spectra, which is less sensitive to temperature structure than emission spectra.  {We also note we chose only one set of infrared and optical coefficients \citep[chosen to most closely match observations presented in ]{Fu2021}. Different choices in these coefficients would lead to slightly different temperature profiles, but this exploration is beyond the scope of this work. Future work will compare the impact of using spectra generated from double gray and picket fence GCMs to determine how robust the patterns identified here are along different radiative transfer schemes and planet parameters.

A physical process absent from these models are clouds. Clouds should reduce the depth of spectral features and flatten the resulting spectra. Additionally, clouds could potentially sculpt the Doppler fields calculated by blocking out particular regions and create the "bottoming out" trend seen after mid transit in \cite{Ehrenreich2020}, as discussed in \cite{savel2022no}.  For a planet of this temperature, one could potentially find some clouds in the nightside upper latitudes \citep[][]{Roman_2021}, so we may initially not expect them to have any strong signatures in the transmission spectra. However, those cloudy models were run in the absence of magnetic effects; when the advection of hot gas to the nightside is reduced, we may expect a colder nightside and therefore more cloud formation. Additionally, work from \cite{Helling2021} suggests that cloud opacity at the morning terminator and ionic or atomic opacity sources at the evening terminator may influence the resulting transmission spectra for UHJs. However, we leave the interplay between magnetic drag and cloud physics for later work. 

Our active magnetic drag 3~G model also makes simplifying assumptions regarding the magnetic field of the planet.  \citep[For a detailed explanation of model assumptions, see][]{Beltz2022a}. The most relevant of these assumptions to this work is that any magnetic field induced in the atmosphere is smaller than the global magnetic field. Mathematically, this results in our prescription being most effective when the magnetic Reynolds number, $R_{m}$, is $ < 1$. This holds true for the vast majority of the planet's atmosphere, but there is a small region in the dayside upper atmosphere where the values of $R_{m}$ reach unity or slightly above. However, given that the dayside is never fully in view during transit, this small region of the atmosphere is likely not very influential in the transmission spectra presented here, but would only influence them secondarily through any change in the day-night circulation. 

\subsection{Combining Observations for Detecting Magnetic Effects}
In this work, we have identified trends in high resolution transmission spectroscopy of planets in the magnetic circulation regime. We perhaps see this trend in the \citet{Ehrenreich2020} data, but to more reliably convince ourselves that this planet (or any other planet) is operating within the magnetic circulation regime, we can combine this trend with others described in  \citet{Beltz2022a} and \citet{Beltz2022b}, therefore allowing our conclusion to become more robust. Combining three independent observations (phase curves, high resolution emission, and high resolution transmission spectroscopy) offers a chance to more conclusively identify planets that are strongly influenced by magnetic effects. We summarize these trends below:
\begin{itemize}

    \item \textit{High Resolution Transmission Spectra:} In this work, we found that for magnetically active models, the net Doppler shift showed less overall blueshifting throughout transit and, depending on the wavelength, could become more redshifted during parts of transit. Neither behavior was shown by the drag free or uniform drag models. 
    \item \textit{High Resolution Emission Spectra:} The magnetic circulation regime influences the net Doppler shift as a function of phase, especially around secondary eclipse for high resolution emission spectra.  Our work in \citet{Beltz2022b} found that our active drag shows a unique trend in net Doppler shift compared to the ones found in the drag free and uniform models near secondary eclipse (see Figure 7 in that paper). 
    \item \textit{Phasecurves: } Our work in \citet{Beltz2022a} found that increasing our magnetic drag strength resulted in a decrease in hotspot offset and an increase in day-night temperature contrast.  
\end{itemize} 

This set of three papers and the trends discussed within can act as roadmap for finding exoplanet atmospheres influenced by magnetic effects.

\section{Conclusion} \label{sec:Conclusion}
In this work, we post-processed three different models of the UHJ WASP-76b with varying forms of magnetic drag treatment to generate high resolution transmission spectra for three different wavelength regimes. The main results of this work are as follows:
\begin{itemize}
    \item 3D effects of both varying temperature and wind structure are present in this high-resolution transmission spectra and  alter the line shape and depth of various features, offering an avenue for assessing sources of drag or magnetic effects within the atmosphere. 
    \item  While transmission spectra from all models generally show increasingly blueshifted net Doppler shifts as transit progresses, the specific patterns and magnitudes depend on the model and wavelength range (and the dominant source of opacity) considered. 
    \item The 3~G model shows the largest differences in Doppler shifts from the other models, beginning with the strongest net redshift of any model as well as actually becoming less blueshifted from phase 0-0.02 for the spectra generated at 1.135-1.355 $\mu$m. This is due to the model possessing the strongest redshifted line of sight velocities during transit, as seen in Figure \ref{fig: losvel}, and may provide a unique way to constrain the role of magnetism within UHJ atmospheres.
    \item Our 3~G model was best able to match the Doppler shift trends in the data presented by \citet{Ehrenreich2020} including the ``bottoming out" behavior during the second half of transit, which only appeared in the kinematic MHD models. Thus, magnetic effects may help explain this particular dataset. 
\end{itemize}

High resolution spectroscopy has opened the door to planetary atmospheric characterization at an unprecedented level, uniquely probing physical processes which were previously unobservable. In order to extract the most meaningful, unbiased conclusions from this data, high complexity atmospheric models and sophisticated post-processing routines are needed in order to account for 3D gradients in temperature, winds, and chemical composition. UHJs, due to their favorable signal to noise ratio, remain the best planetary target for investigating analysis techniques for this type of data. However, these planets have the largest spatial gradients and, due to their high temperatures, must have partially ionized atmospheres. It is therefore necessary to consider how magnetic effects may shape the spectra of these planets and, in turn, how those spectra can give us insight into the physical states of the atmospheres.  

\section{Acknowledgments}
This work was generously supported by a grant from the Heising-Simons Foundation. Many of the calculations in this paper made use of the Great Lakes High Performance Computing Cluster maintained by the University of Michigan. We thank the reviewer for their comments, which improved the quality of this manuscript.

\bibliographystyle{aasjournal}
\bibliography{bib.bib}

\begin{thebibliography}{}
\expandafter\ifx\csname natexlab\endcsname\relax\def\natexlab#1{#1}\fi
\providecommand{\url}[1]{\href{#1}{#1}}
\providecommand{\dodoi}[1]{doi:~\href{http://doi.org/#1}{\nolinkurl{#1}}}
\providecommand{\doeprint}[1]{\href{http://ascl.net/#1}{\nolinkurl{http://ascl.net/#1}}}
\providecommand{\doarXiv}[1]{\href{https://arxiv.org/abs/#1}{\nolinkurl{https://arxiv.org/abs/#1}}}

\bibitem[{Bell \& Cowan(2018)}]{Bell_2018}
Bell, T.~J., \& Cowan, N.~B. 2018, The Astrophysical Journal, 857, L20,
  \dodoi{10.3847/2041-8213/aabcc8}

\bibitem[{{Beltz} {et~al.}(2021){Beltz}, {Rauscher}, {Brogi}, \&
  {Kempton}}]{Beltz2021}
{Beltz}, H., {Rauscher}, E., {Brogi}, M., \& {Kempton}, E. M.~R. 2021, \aj,
  161, 1, \dodoi{10.3847/1538-3881/abb67b}

\bibitem[{{Beltz} {et~al.}(2022{\natexlab{a}}){Beltz}, {Rauscher}, {Kempton},
  {Malsky}, {Ochs}, {Arora}, \& {Savel}}]{Beltz2022b}
{Beltz}, H., {Rauscher}, E., {Kempton}, E. M.~R., {et~al.} 2022{\natexlab{a}},
  \aj, 164, 140, \dodoi{10.3847/1538-3881/ac897b}

\bibitem[{{Beltz} {et~al.}(2022{\natexlab{b}}){Beltz}, {Rauscher}, {Roman}, \&
  {Guilliat}}]{Beltz2022a}
{Beltz}, H., {Rauscher}, E., {Roman}, M.~T., \& {Guilliat}, A.
  2022{\natexlab{b}}, \aj, 163, 35, \dodoi{10.3847/1538-3881/ac3746}

\bibitem[{{Brogi} \& {Birkby}(2021)}]{Brogibook2021}
{Brogi}, M., \& {Birkby}, J. 2021, in ExoFrontiers; Big Questions in
  Exoplanetary Science, ed. N.~{Madhusudhan}, 8--1,
  \dodoi{10.1088/2514-3433/abfa8fch8}

\bibitem[{{Brogi} {et~al.}(2016){Brogi}, {de Kok}, {Albrecht}, {Snellen},
  {Birkby}, \& {Schwarz}}]{Brogi2016}
{Brogi}, M., {de Kok}, R.~J., {Albrecht}, S., {et~al.} 2016, \apj, 817, 106,
  \dodoi{10.3847/0004-637X/817/2/106}

\bibitem[{{Carone} {et~al.}(2020){Carone}, {Baeyens}, {Molli{\`e}re}, {Barth},
  {Vazan}, {Decin}, {Sarkis}, {Venot}, \& {Henning}}]{Carone2020}
{Carone}, L., {Baeyens}, R., {Molli{\`e}re}, P., {et~al.} 2020, \mnras, 496,
  3582, \dodoi{10.1093/mnras/staa1733}

\bibitem[{{Casasayas-Barris} {et~al.}(2021){Casasayas-Barris}, {Orell-Miquel},
  {Stangret}, {Nortmann}, {Yan}, {Oshagh}, {Palle}, {Sanz-Forcada},
  {L{\'o}pez-Puertas}, {Nagel}, {Luque}, {Morello}, {Snellen}, {Zechmeister},
  {Quirrenbach}, {Caballero}, {Ribas}, {Reiners}, {Amado}, {Bergond}, {Czesla},
  {Henning}, {Khalafinejad}, {Molaverdikhani}, {Montes}, {Perger},
  {S{\'a}nchez-L{\'o}pez}, \& {Sedaghati}}]{Cassayas2021}
{Casasayas-Barris}, N., {Orell-Miquel}, J., {Stangret}, M., {et~al.} 2021,
  \aap, 654, A163, \dodoi{10.1051/0004-6361/202141669}

\bibitem[{{Deibert} {et~al.}(2021){Deibert}, {de Mooij}, {Jayawardhana},
  {Turner}, {Ridden-Harper}, {Fossati}, {Hood}, {Fortney}, {Flagg},
  {MacDonald}, {Allart}, \& {Sing}}]{Deibert2021}
{Deibert}, E.~K., {de Mooij}, E. J.~W., {Jayawardhana}, R., {et~al.} 2021,
  \apjl, 919, L15, \dodoi{10.3847/2041-8213/ac2513}

\bibitem[{Deitrick {et~al.}(2020)Deitrick, Mendon{\c{c}}a, Schroffenegger,
  Grimm, Tsai, \& Heng}]{Deitrick_2020THOR}
Deitrick, R., Mendon{\c{c}}a, J.~M., Schroffenegger, U., {et~al.} 2020, The
  Astrophysical Journal Supplement Series, 248, 30,
  \dodoi{10.3847/1538-4365/ab930e}

\bibitem[{{Ehrenreich} {et~al.}(2020){Ehrenreich}, {Lovis}, {Allart}, {Zapatero
  Osorio}, {Pepe}, {Cristiani}, {Rebolo}, {Santos}, {Borsa}, {Demangeon},
  {Dumusque}, {Gonz{\'a}lez Hern{\'a}ndez}, {Casasayas-Barris},
  {S{\'e}gransan}, {Sousa}, {Abreu}, {Adibekyan}, {Affolter}, {Allende Prieto},
  {Alibert}, {Aliverti}, {Alves}, {Amate}, {Avila}, {Baldini}, {Bandy}, {Benz},
  {Bianco}, {Bolmont}, {Bouchy}, {Bourrier}, {Broeg}, {Cabral}, {Calderone},
  {Pall{\'e}}, {Cegla}, {Cirami}, {Coelho}, {Conconi}, {Coretti}, {Cumani},
  {Cupani}, {Dekker}, {Delabre}, {Deiries}, {D'Odorico}, {Di Marcantonio},
  {Figueira}, {Fragoso}, {Genolet}, {Genoni}, {G{\'e}nova Santos}, {Hara},
  {Hughes}, {Iwert}, {Kerber}, {Knudstrup}, {Landoni}, {Lavie}, {Lizon},
  {Lendl}, {Lo Curto}, {Maire}, {Manescau}, {Martins}, {M{\'e}gevand},
  {Mehner}, {Micela}, {Modigliani}, {Molaro}, {Monteiro}, {Monteiro},
  {Moschetti}, {M{\"u}ller}, {Nunes}, {Oggioni}, {Oliveira}, {Pariani},
  {Pasquini}, {Poretti}, {Rasilla}, {Redaelli}, {Riva}, {Santana Tschudi},
  {Santin}, {Santos}, {Segovia Milla}, {Seidel}, {Sosnowska}, {Sozzetti},
  {Span{\`o}}, {Su{\'a}rez Mascare{\~n}o}, {Tabernero}, {Tenegi}, {Udry},
  {Zanutta}, \& {Zerbi}}]{Ehrenreich2020}
{Ehrenreich}, D., {Lovis}, C., {Allart}, R., {et~al.} 2020, \nat, 580, 597,
  \dodoi{10.1038/s41586-020-2107-1}

\bibitem[{{Flowers} {et~al.}(2019){Flowers}, {Brogi}, {Rauscher}, {Kempton}, \&
  {Chiavassa}}]{Flowers2019}
{Flowers}, E., {Brogi}, M., {Rauscher}, E., {Kempton}, E. M.~R., \&
  {Chiavassa}, A. 2019, \aj, 157, 209, \dodoi{10.3847/1538-3881/ab164c}

\bibitem[{{Follert} {et~al.}(2014){Follert}, {Dorn}, {Oliva}, {Lizon},
  {Hatzes}, {Piskunov}, {Reiners}, {Seemann}, {Stempels}, {Heiter}, {Marquart},
  {Lockhart}, {Anglada-Escude}, {L{\"o}winger}, {Baade}, {Grunhut}, {Bristow},
  {Klein}, {Jung}, {Ives}, {Kerber}, {Pozna}, {Paufique}, {Kaeufl}, {Origlia},
  {Valenti}, {Gojak}, {Hilker}, {Pasquini}, {Smette}, \&
  {Smoker}}]{CRIRES+2014}
{Follert}, R., {Dorn}, R.~J., {Oliva}, E., {et~al.} 2014, in Society of
  Photo-Optical Instrumentation Engineers (SPIE) Conference Series, Vol. 9147,
  Ground-based and Airborne Instrumentation for Astronomy V, ed. S.~K.
  {Ramsay}, I.~S. {McLean}, \& H.~{Takami}, 914719, \dodoi{10.1117/12.2054197}

\bibitem[{{Fu} {et~al.}(2021){Fu}, {Deming}, {Lothringer}, {Nikolov}, {Sing},
  {Kempton}, {Ih}, {Evans}, {Stevenson}, {Wakeford}, {Rodriguez}, {Eastman},
  {Stassun}, {Henry}, {L{\'o}pez-Morales}, {Lendl}, {Conti}, {Stockdale},
  {Collins}, {Kielkopf}, {Barstow}, {Sanz-Forcada}, {Ehrenreich}, {Bourrier},
  \& {dos Santos}}]{Fu2021}
{Fu}, G., {Deming}, D., {Lothringer}, J., {et~al.} 2021, \aj, 162, 108,
  \dodoi{10.3847/1538-3881/ac1200}

\bibitem[{{Gandhi} {et~al.}(2022){Gandhi}, {Kesseli}, {Snellen}, {Brogi},
  {Wardenier}, {Parmentier}, {Welbanks}, \& {Savel}}]{Gandhi2022}
{Gandhi}, S., {Kesseli}, A., {Snellen}, I., {et~al.} 2022, \mnras, 515, 749,
  \dodoi{10.1093/mnras/stac1744}

\bibitem[{{Helling} {et~al.}(2021){Helling}, {Lewis}, {Samra}, {Carone},
  {Graham}, {Herbort}, {Chubb}, {Min}, {Waters}, {Parmentier}, \&
  {Mayne}}]{Helling2021}
{Helling}, C., {Lewis}, D., {Samra}, D., {et~al.} 2021, \aap, 649, A44,
  \dodoi{10.1051/0004-6361/202039911}

\bibitem[{{Herman} {et~al.}(2022){Herman}, {de Mooij}, {Nugroho}, {Gibson}, \&
  {Jayawardhana}}]{Herman2022}
{Herman}, M.~K., {de Mooij}, E. J.~W., {Nugroho}, S.~K., {Gibson}, N.~P., \&
  {Jayawardhana}, R. 2022, \aj, 163, 248, \dodoi{10.3847/1538-3881/ac5f4d}

\bibitem[{{Ikeda} {et~al.}(2016){Ikeda}, {Kobayashi}, {Kondo}, {Otsubo},
  {Hamano}, {Sameshima}, {Yoshikawa}, {Fukue}, {Nakanishi}, {Kawanishi},
  {Nakaoka}, {Kinoshita}, {Kitano}, {Asano}, {Takenaka}, {Watase}, {Mito},
  {Yasui}, {Minami}, {Izumu}, {Yamamoto}, {Mizumoto}, {Arasaki}, {Arai},
  {Matsunaga}, \& {Kawakita}}]{WINERED2016}
{Ikeda}, Y., {Kobayashi}, N., {Kondo}, S., {et~al.} 2016, in Society of
  Photo-Optical Instrumentation Engineers (SPIE) Conference Series, Vol. 9908,
  Ground-based and Airborne Instrumentation for Astronomy VI, ed. C.~J.
  {Evans}, L.~{Simard}, \& H.~{Takami}, 99085Z, \dodoi{10.1117/12.2230886}

\bibitem[{{Kesseli} {et~al.}(2022){Kesseli}, {Snellen}, {Casasayas-Barris},
  {Molli{\`e}re}, \& {S{\'a}nchez-L{\'o}pez}}]{Kesseli2022w76spectralsurvey}
{Kesseli}, A.~Y., {Snellen}, I.~A.~G., {Casasayas-Barris}, N., {Molli{\`e}re},
  P., \& {S{\'a}nchez-L{\'o}pez}, A. 2022, \aj, 163, 107,
  \dodoi{10.3847/1538-3881/ac4336}

\bibitem[{{Kreidberg} {et~al.}(2018){Kreidberg}, {Line}, {Parmentier},
  {Stevenson}, {Louden}, {Bonnefoy}, {Faherty}, {Henry}, {Williamson},
  {Stassun}, {Beatty}, {Bean}, {Fortney}, {Showman}, {D{\'e}sert}, \&
  {Arcangeli}}]{Kreidberg2018}
{Kreidberg}, L., {Line}, M.~R., {Parmentier}, V., {et~al.} 2018, \aj, 156, 17,
  \dodoi{10.3847/1538-3881/aac3df}

\bibitem[{Kurucz(1995)}]{kurucz1995kurucz}
Kurucz, R. 1995, Atomic Line List

\bibitem[{{Landman} {et~al.}(2021){Landman}, {S{\'a}nchez-L{\'o}pez},
  {Molli{\`e}re}, {Kesseli}, {Louca}, \& {Snellen}}]{Landman2021}
{Landman}, R., {S{\'a}nchez-L{\'o}pez}, A., {Molli{\`e}re}, P., {et~al.} 2021,
  \aap, 656, A119, \dodoi{10.1051/0004-6361/202141696}

\bibitem[{{Lee} {et~al.}(2022){Lee}, {Prinoth}, {Kitzmann}, {Tsai},
  {Hoeijmakers}, {Borsato}, \& {Heng}}]{Lee2022}
{Lee}, E. K.~H., {Prinoth}, B., {Kitzmann}, D., {et~al.} 2022, \mnras, 517,
  240, \dodoi{10.1093/mnras/stac2246}

\bibitem[{{Lee} {et~al.}(2021){Lee}, {Wardenier}, {Prinoth}, {Parmentier},
  {Grimm}, {Baeyens}, {Carone}, {Christie}, {Deitrick}, {Kitzmann}, {Mayne}, \&
  {Roman}}]{Lee2021}
{Lee}, E. K.~H., {Wardenier}, J.~P., {Prinoth}, B., {et~al.} 2021, arXiv
  e-prints, arXiv:2110.15640.
\newblock \doarXiv{2110.15640}

\bibitem[{{Lodders}(2003)}]{Lodders2003}
{Lodders}, K. 2003, \apj, 591, 1220, \dodoi{10.1086/375492}

\bibitem[{{Louden} \& {Wheatley}(2015)}]{Louden2015}
{Louden}, T., \& {Wheatley}, P.~J. 2015, \apjl, 814, L24,
  \dodoi{10.1088/2041-8205/814/2/L24}

\bibitem[{{May} {et~al.}(2021){May}, {Komacek}, {Stevenson}, {Kempton}, {Bean},
  {Malik}, {Ih}, {Mansfield}, {Savel}, {Deming}, {Desert}, {Feng}, {Fortney},
  {Kataria}, {Lewis}, {Morley}, {Rauscher}, \& {Showman}}]{May2021}
{May}, E.~M., {Komacek}, T.~D., {Stevenson}, K.~B., {et~al.} 2021, \aj, 162,
  158, \dodoi{10.3847/1538-3881/ac0e30}

\bibitem[{Menou(2012)}]{Menou_2012}
Menou, K. 2012, The Astrophysical Journal, 745, 138,
  \dodoi{10.1088/0004-637x/745/2/138}

\bibitem[{{Miller-Ricci Kempton} \& {Rauscher}(2012)}]{Kempton2012}
{Miller-Ricci Kempton}, E., \& {Rauscher}, E. 2012, \apj, 751, 117,
  \dodoi{10.1088/0004-637X/751/2/117}

\bibitem[{{Park} {et~al.}(2014){Park}, {Jaffe}, {Yuk}, {Chun}, {Pak}, {Kim},
  {Pavel}, {Lee}, {Oh}, {Jeong}, {Sim}, {Lee}, {Nguyen Le}, {Strubhar},
  {Gully-Santiago}, {Oh}, {Cha}, {Moon}, {Park}, {Brooks}, {Ko}, {Han}, {Nah},
  {Hill}, {Lee}, {Barnes}, {Yu}, {Kaplan}, {Mace}, {Kim}, {Lee}, {Hwang}, \&
  {Park}}]{IGRINS2014}
{Park}, C., {Jaffe}, D.~T., {Yuk}, I.-S., {et~al.} 2014, in Society of
  Photo-Optical Instrumentation Engineers (SPIE) Conference Series, Vol. 9147,
  Ground-based and Airborne Instrumentation for Astronomy V, ed. S.~K.
  {Ramsay}, I.~S. {McLean}, \& H.~{Takami}, 91471D, \dodoi{10.1117/12.2056431}

\bibitem[{{Parmentier} {et~al.}(2018){Parmentier}, {Line}, {Bean}, {Mansfield},
  {Kreidberg}, {Lupu}, {Visscher}, {D{\'e}sert}, {Fortney}, {Deleuil},
  {Arcangeli}, {Showman}, \& {Marley}}]{parmentier2018}
{Parmentier}, V., {Line}, M.~R., {Bean}, J.~L., {et~al.} 2018, \aap, 617, A110,
  \dodoi{10.1051/0004-6361/201833059}

\bibitem[{{Perna} {et~al.}(2010){Perna}, {Menou}, \&
  {Rauscher}}]{Perna2010magdrag}
{Perna}, R., {Menou}, K., \& {Rauscher}, E. 2010, \apj, 719, 1421,
  \dodoi{10.1088/0004-637X/719/2/1421}

\bibitem[{{Pino} {et~al.}(2022){Pino}, {Brogi}, {D{\'e}sert}, {Nascimbeni},
  {Bonomo}, {Rauscher}, {Basilicata}, {Biazzo}, {Bignamini}, {Borsa}, {Claudi},
  {Covino}, {Di Mauro}, {Guilluy}, {Maggio}, {Malavolta}, {Micela}, {Molinari},
  {Molinaro}, {Montalto}, {Nardiello}, {Pedani}, {Piotto}, {Poretti}, {Rainer},
  {Scandariato}, {Sicilia}, \& {Sozzetti}}]{Pino2022}
{Pino}, L., {Brogi}, M., {D{\'e}sert}, J.~M., {et~al.} 2022, arXiv e-prints,
  arXiv:2209.11735.
\newblock \doarXiv{2209.11735}

\bibitem[{{Pluriel} {et~al.}(2020){Pluriel}, {Zingales}, {Leconte}, \&
  {Parmentier}}]{Pluriel2020}
{Pluriel}, W., {Zingales}, T., {Leconte}, J., \& {Parmentier}, V. 2020, \aap,
  636, A66, \dodoi{10.1051/0004-6361/202037678}

\bibitem[{Polyansky {et~al.}(2018)Polyansky, Kyuberis, Zobov, Tennyson,
  Yurchenko, \& Lodi}]{polyansky2018exomol}
Polyansky, O.~L., Kyuberis, A.~A., Zobov, N.~F., {et~al.} 2018, Monthly Notices
  of the Royal Astronomical Society, 480, 2597

\bibitem[{{Quirrenbach} {et~al.}(2014){Quirrenbach}, {Amado}, {Caballero},
  {Mundt}, {Reiners}, {Ribas}, {Seifert}, {Abril}, {Aceituno},
  {Alonso-Floriano}, {Ammler-von Eiff}, {Antona Jim{\'e}nez},
  {Anwand-Heerwart}, {Azzaro}, {Bauer}, {Barrado}, {Becerril}, {B{\'e}jar},
  {Ben{\'\i}tez}, {Berdi{\~n}as}, {C{\'a}rdenas}, {Casal}, {Claret},
  {Colom{\'e}}, {Cort{\'e}s-Contreras}, {Czesla}, {Doellinger}, {Dreizler},
  {Feiz}, {Fern{\'a}ndez}, {Galad{\'\i}}, {G{\'a}lvez-Ortiz},
  {Garc{\'\i}a-Piquer}, {Garc{\'\i}a-Vargas}, {Garrido}, {Gesa}, {G{\'o}mez
  Galera}, {Gonz{\'a}lez {\'A}lvarez}, {Gonz{\'a}lez Hern{\'a}ndez},
  {Gr{\"o}zinger}, {Gu{\`a}rdia}, {Guenther}, {de Guindos},
  {Guti{\'e}rrez-Soto}, {Hagen}, {Hatzes}, {Hauschildt}, {Helmling}, {Henning},
  {Hermann}, {Hern{\'a}ndez Casta{\~n}o}, {Herrero}, {Hidalgo}, {Holgado},
  {Huber}, {Huber}, {Jeffers}, {Joergens}, {de Juan}, {Kehr}, {Klein},
  {K{\"u}rster}, {Lamert}, {Lalitha}, {Laun}, {Lemke}, {Lenzen}, {L{\'o}pez del
  Fresno}, {L{\'o}pez Mart{\'\i}}, {L{\'o}pez-Santiago}, {Mall}, {Mandel},
  {Mart{\'\i}n}, {Mart{\'\i}n-Ruiz}, {Mart{\'\i}nez-Rodr{\'\i}guez}, {Marvin},
  {Mathar}, {Mirabet}, {Montes}, {Morales Mu{\~n}oz}, {Moya}, {Naranjo},
  {Ofir}, {Oreiro}, {Pall{\'e}}, {Panduro}, {Passegger}, {P{\'e}rez-Calpena},
  {P{\'e}rez Medialdea}, {Perger}, {Pluto}, {Ram{\'o}n}, {Rebolo}, {Redondo},
  {Reffert}, {Reinhardt}, {Rhode}, {Rix}, {Rodler}, {Rodr{\'\i}guez},
  {Rodr{\'\i}guez-L{\'o}pez}, {Rodr{\'\i}guez-P{\'e}rez}, {Rohloff}, {Rosich},
  {S{\'a}nchez-Blanco}, {S{\'a}nchez Carrasco}, {Sanz-Forcada}, {Sarmiento},
  {Sch{\"a}fer}, {Schiller}, {Schmidt}, {Schmitt}, {Solano}, {Stahl}, {Storz},
  {St{\"u}rmer}, {Su{\'a}rez}, {Ulbrich}, {Veredas}, {Wagner}, {Winkler},
  {Zapatero Osorio}, {Zechmeister}, {Abell{\'a}n de Paco},
  {Anglada-Escud{\'e}}, {del Burgo}, {Klutsch}, {Lizon}, {L{\'o}pez-Morales},
  {Morales}, {Perryman}, {Tulloch}, \& {Xu}}]{CARMENES2014}
{Quirrenbach}, A., {Amado}, P.~J., {Caballero}, J.~A., {et~al.} 2014, in
  Society of Photo-Optical Instrumentation Engineers (SPIE) Conference Series,
  Vol. 9147, Ground-based and Airborne Instrumentation for Astronomy V, ed.
  S.~K. {Ramsay}, I.~S. {McLean}, \& H.~{Takami}, 91471F,
  \dodoi{10.1117/12.2056453}

\bibitem[{Rauscher \& Menou(2012)}]{Rauscher2012GCM}
Rauscher, E., \& Menou, K. 2012, The Astrophysical Journal, 750, 96.
\newblock \url{http://stacks.iop.org/0004-637X/750/i=2/a=96}

\bibitem[{Rauscher \& Menou(2013)}]{RauscherMenou2013}
---. 2013, The Astrophysical Journal, 764, 103,
  \dodoi{10.1088/0004-637x/764/1/103}

\bibitem[{{Rogers}(2017)}]{Rogers2017}
{Rogers}, T.~M. 2017, Nature Astronomy, 1, 0131,
  \dodoi{10.1038/s41550-017-0131}

\bibitem[{Rogers \& Komacek(2014)}]{Rogers_2014b}
Rogers, T.~M., \& Komacek, T.~D. 2014, The Astrophysical Journal, 794, 132,
  \dodoi{10.1088/0004-637x/794/2/132}

\bibitem[{Rogers \& Showman(2014)}]{Rogers_2014_showman}
Rogers, T.~M., \& Showman, A.~P. 2014, The Astrophysical Journal, 782, L4,
  \dodoi{10.1088/2041-8205/782/1/l4}

\bibitem[{Roman \& Rauscher(2017)}]{newradRomanRausher}
Roman, M., \& Rauscher, E. 2017, The Astrophysical Journal, 850, 17.
\newblock \url{http://stacks.iop.org/0004-637X/850/i=1/a=17}

\bibitem[{Roman {et~al.}(2021)Roman, Kempton, Rauscher, Harada, Bean, \&
  Stevenson}]{Roman_2021}
Roman, M.~T., Kempton, E. M.-R., Rauscher, E., {et~al.} 2021, The Astrophysical
  Journal, 908, 101, \dodoi{10.3847/1538-4357/abd549}

\bibitem[{Rothman {et~al.}(2010)Rothman, Gordon, Barber, Dothe, Gamache,
  Goldman, Perevalov, Tashkun, \& Tennyson}]{rothman2010hitemp}
Rothman, L.~S., Gordon, I., Barber, R., {et~al.} 2010, Journal of Quantitative
  Spectroscopy and Radiative Transfer, 111, 2139

\bibitem[{{Savel} {et~al.}(2023){Savel}, {Kempton}, {Rauscher}, {Komacek},
  {Bean}, {Malik}, \& {Malsky}}]{Savel2023}
{Savel}, A.~B., {Kempton}, E. M.~R., {Rauscher}, E., {et~al.} 2023, arXiv
  e-prints, arXiv:2301.01694.
\newblock \doarXiv{2301.01694}

\bibitem[{Savel {et~al.}(2022)Savel, Kempton, Malik, Komacek, Bean, May,
  Stevenson, Mansfield, \& Rauscher}]{savel2022no}
Savel, A.~B., Kempton, E. M.-R., Malik, M., {et~al.} 2022, The Astrophysical
  Journal, 926, 85

\bibitem[{{Schwarz} {et~al.}(2016){Schwarz}, {Ginski}, {de Kok}, {Snellen},
  {Brogi}, \& {Birkby}}]{Schwarz2016}
{Schwarz}, H., {Ginski}, C., {de Kok}, R.~J., {et~al.} 2016, \aap, 593, A74,
  \dodoi{10.1051/0004-6361/201628908}

\bibitem[{{Snellen} {et~al.}(2010){Snellen}, {de Kok}, {de Mooij}, \&
  {Albrecht}}]{Snellen2010}
{Snellen}, I. A.~G., {de Kok}, R.~J., {de Mooij}, E. J.~W., \& {Albrecht}, S.
  2010, \nat, 465, 1049, \dodoi{10.1038/nature09111}

\bibitem[{{Stock} {et~al.}(2022){Stock}, {Kitzmann}, \& {Patzer}}]{Stock2018}
{Stock}, J.~W., {Kitzmann}, D., \& {Patzer}, A. B.~C. 2022, \mnras, 517, 4070,
  \dodoi{10.1093/mnras/stac2623}

\bibitem[{{Stock} {et~al.}(2018){Stock}, {Kitzmann}, {Patzer}, \&
  {Sedlmayr}}]{Fastchem2018}
{Stock}, J.~W., {Kitzmann}, D., {Patzer}, A. B.~C., \& {Sedlmayr}, E. 2018,
  \mnras, 479, 865, \dodoi{10.1093/mnras/sty1531}

\bibitem[{{Tabernero} {et~al.}(2021){Tabernero}, {Zapatero Osorio}, {Allart},
  {Borsa}, {Casasayas-Barris}, {Demangeon}, {Ehrenreich}, {Lillo-Box}, {Lovis},
  {Pall{\'e}}, {Sousa}, {Rebolo}, {Santos}, {Pepe}, {Cristiani}, {Adibekyan},
  {Allende Prieto}, {Alibert}, {Barros}, {Bouchy}, {Bourrier}, {D'Odorico},
  {Dumusque}, {Faria}, {Figueira}, {G{\'e}nova Santos}, {Gonz{\'a}lez
  Hern{\'a}ndez}, {Hojjatpanah}, {Lo Curto}, {Lavie}, {Martins}, {Martins},
  {Mehner}, {Micela}, {Molaro}, {Nunes}, {Poretti}, {Seidel}, {Sozzetti},
  {Su{\'a}rez Mascare{\~n}o}, {Udry}, {Aliverti}, {Affolter}, {Alves}, {Amate},
  {Avila}, {Bandy}, {Benz}, {Bianco}, {Broeg}, {Cabral}, {Conconi}, {Coelho},
  {Cumani}, {Deiries}, {Dekker}, {Delabre}, {Fragoso}, {Genoni}, {Genolet},
  {Hughes}, {Knudstrup}, {Kerber}, {Landoni}, {Lizon}, {Maire}, {Manescau}, {Di
  Marcantonio}, {M{\'e}gevand}, {Monteiro}, {Monteiro}, {Moschetti}, {Mueller},
  {Modigliani}, {Oggioni}, {Oliveira}, {Pariani}, {Pasquini}, {Rasilla},
  {Redaelli}, {Riva}, {Santana-Tschudi}, {Santin}, {Santos}, {Segovia},
  {Sosnowska}, {Span{\`o}}, {Tenegi}, {Iwert}, {Zanutta}, \&
  {Zerbi}}]{Taberno2021}
{Tabernero}, H.~M., {Zapatero Osorio}, M.~R., {Allart}, R., {et~al.} 2021,
  \aap, 646, A158, \dodoi{10.1051/0004-6361/202039511}

\bibitem[{Tan \& Komacek(2019)}]{Tan_2019}
Tan, X., \& Komacek, T.~D. 2019, The Astrophysical Journal, 886, 26,
  \dodoi{10.3847/1538-4357/ab4a76}

\bibitem[{{van Sluijs} {et~al.}(2022){van Sluijs}, {Birkby}, {Lothringer},
  {Lee}, {Crossfield}, {Parmentier}, {Brogi}, {Kulesa}, {McCarthy}, {Powell},
  \& {Charbonneau}}]{vansluijs2022}
{van Sluijs}, L., {Birkby}, J.~L., {Lothringer}, J., {et~al.} 2022, arXiv
  e-prints, arXiv:2203.13234.
\newblock \doarXiv{2203.13234}

\bibitem[{{Wardenier} {et~al.}(2022){Wardenier}, {Parmentier}, \&
  {Lee}}]{Wardenier2022}
{Wardenier}, J.~P., {Parmentier}, V., \& {Lee}, E. K.~H. 2022, \mnras, 510,
  620, \dodoi{10.1093/mnras/stab3432}

\bibitem[{{Wardenier} {et~al.}(2021){Wardenier}, {Parmentier}, {Lee}, {Line},
  \& {Gharib-Nezhad}}]{Wardenier2021}
{Wardenier}, J.~P., {Parmentier}, V., {Lee}, E. K.~H., {Line}, M.~R., \&
  {Gharib-Nezhad}, E. 2021, \mnras, 506, 1258, \dodoi{10.1093/mnras/stab1797}

\bibitem[{{West} {et~al.}(2016){West}, {Hellier}, {Almenara}, {Anderson},
  {Barros}, {Bouchy}, {Brown}, {Collier Cameron}, {Deleuil}, {Delrez}, {Doyle},
  {Faedi}, {Fumel}, {Gillon}, {G{\'o}mez Maqueo Chew}, {H{\'e}brard}, {Jehin},
  {Lendl}, {Maxted}, {Pepe}, {Pollacco}, {Queloz}, {S{\'e}gransan}, {Smalley},
  {Smith}, {Southworth}, {Triaud}, \& {Udry}}]{West2016}
{West}, R.~G., {Hellier}, C., {Almenara}, J.~M., {et~al.} 2016, \aap, 585,
  A126, \dodoi{10.1051/0004-6361/201527276}

\bibitem[{{Yadav} \& {Thorngren}(2017)}]{Yadav2017}
{Yadav}, R.~K., \& {Thorngren}, D.~P. 2017, \apjl, 849, L12,
  \dodoi{10.3847/2041-8213/aa93fd}

\bibitem[{{Zhang} {et~al.}(2022){Zhang}, {Snellen}, {Wyttenbach}, {Nielsen},
  {Lendl}, {Casasayas-Barris}, {Chaverot}, {Kesseli}, {Lovis}, {Pepe},
  {Psaridi}, {Seidel}, {Udry}, \& {Ulmer-Moll}}]{Zhang2022Hydrostatic}
{Zhang}, Y., {Snellen}, I. A.~G., {Wyttenbach}, A., {et~al.} 2022, \aap, 666,
  A47, \dodoi{10.1051/0004-6361/202244203}

\end{thebibliography}

\end{document}